\documentclass[prl,twocolumn,showpacs,floatfix,superscriptaddress]{revtex4}

\usepackage{amssymb}

\usepackage{graphicx}
\usepackage{dcolumn}
\usepackage{bm}
\usepackage{color}
\begin{document}

\title{Anatomy of perpendicular magnetic anisotropy in Fe/MgO magnetic tunnel junctions: First principles insight}
\date{\today}
\author{A.~Hallal}
\affiliation{SPINTEC, UMR 8191 CEA-INAC$|$CNRS$|$UJF-Grenoble 1$|$Grenoble-INP, Grenoble, 38054, France}

\author{H.~X.~Yang}
\affiliation{SPINTEC, UMR 8191 CEA-INAC$|$CNRS$|$UJF-Grenoble 1$|$Grenoble-INP, Grenoble, 38054, France}

\author{B.~Dieny}
\affiliation{SPINTEC, UMR 8191 CEA-INAC$|$CNRS$|$UJF-Grenoble 1$|$Grenoble-INP, Grenoble, 38054, France}

\author{M.~Chshiev}
\affiliation{SPINTEC, UMR 8191 CEA-INAC$|$CNRS$|$UJF-Grenoble 1$|$Grenoble-INP, Grenoble, 38054, France}

\begin{abstract}

Using first-principles calculations, we elucidate microscopic mechanisms of perpendicular magnetic anisotropy (PMA)in Fe/MgO magnetic tunnel
junctions through evaluation of orbital and layer resolved contributions into the total anisotropy value. It is demonstrated that the origin of the large PMA values is far beyond simply considering the hybridization between Fe-$3d$ and O-$2p$ orbitals at the interface between the metal and the insulator. On-site projected analysis show that the anisotropy energy is not localized at the interface but it rather propagates into the bulk showing an attenuating oscillatory behavior which depends on orbital character of contributing states and interfacial conditions. Furthermore, it is found in most situations that states with $d_{yz(xz)}$ and $d_{z^2}$ character tend always to maintain the PMA while those with $d_{xy}$ and $d_{x^2-y^2}$ character tend to favor the in-plane anisotropy. It is also found that while MgO thickness has no influence on PMA, the calculated perpendicular magnetic anisotropy oscillates as a function of Fe thickness with a period of 2ML and reaches a maximum value of 3.6~mJ/m$^2$.

\end{abstract}
\maketitle

Perpendicular magnetic anisotropy (PMA) at ferromagnetic transition metal/insulator interfaces has become of huge interest in the context of development of various spintronic devices based on spin-transfer torque or spin-orbit torque (spin-hall or Rashba effect). In particular, out-of-plane magnetized magnetic tunnel junctions (pMTJ) are now intensively developed for spin transfer torque (STT) magnetic random access memories (STT-MRAM) applications where the strong perpendicular anisotropy originating from the CoFe/MgO interface allows to maintain the thermal stability of the storage layer magnetization down to at least the 20nm technological node ~\cite{pMTJ1,pMTJ2,pMTJ3,pMTJ4,pMTJ5,Lavinia,LaviniaPRB,Ohno,MgOPMAAPLohno}.
This interest is due to the fact that it makes possible to avoid introducing within or next to the ferromagnet heavy non-magnetic elements (in particular Pt, Pd, Au etc)
which were believed to be essential to trigger the PMA thanks to their large spin-orbit coupling (SOC)
~\cite{Rashba,reviewRashba,Pd1,Pd2,weller}. However, introducing these heavy elements is detrimental for STT based devices  since their large spin-orbit coupling tends to
increase the Gilbert damping resulting in an increase in the critical current required for switching the storage layer magnetization by STT. This interfacial PMA at CoFe/MgO interface is remarkably large despite the weak SOC. Indeed, PMA with large values up to 1 to 2 mJ/m$^2$ have been reported at Co(Fe)/MOx interfaces (M=Ta, Mg, Al, Ru etc)~\cite{monso,rodmacq1,Lavinia,Suzuki}. These values are comparable to those observed at Co/Pt interface which is considered as a reference for large interfacial anisotropy ~\cite{weller}. Experimentally, it was observed by X-Ray photoemission (XPS) and X-Ray absorption experiments ~\cite{manchon} that the interfacial PMA at ferromagnetic transition metal/oxide gets maximum when oxygen is present along the metal/oxide interface so that chemical bounds (hybridization) can form between the metallic ions orbitals and the oxygen orbitals. Under or over-oxidized interfaces yield weaker PMA ~\cite{monso,rodmacq1,Lavinia,manchon}.
Because of this remarkable combination of large anisotropy and weak SOC, this phenomenon is now widely used in pMTJs for high density STT-MRAM~\cite{Ohno,MgOPMAAPLohno}. 

This phenomenon attracted a large attention from theoretical point of view. Using first-principles calculations, several groups addressed magnetic 
anisotropy in Fe/MgO interfaces and reported values between 1 and 2 mJ/m$^2$ for pure Fe/MgO interfaces~\cite{TsymbalPMA,Nakamura,MgOPMA} as well 
as a surprisingly large PMA value of 19.5~mJ/m$^2$~\cite{GiantPMA}. It was also found that in over- or under-oxidized interfaces, the PMA values 
decrease~\cite{TsymbalPMA,nakajima,GiantPMA,MgOPMA}, which is in a good agreement with experimental observations~\cite{monso, rodmacq1,LaviniaIEEE, APL2012,manchon}. 

Although PMA has been extensively studied both experimentally and theoretically, the origin of its very large value with such weak SOI system has not yet been
fully unveiled. The large PMA in (Co)Fe/MgO(AlOx) is usually interpreted in terms of strong hybridizations between interfacial (Co)Fe-3$d$ and the O-2$p$
and orbitals combined with spin-orbit~\cite{rodmacq,Ohno,MgOPMA}. This interpretation leads to a picture in which the magnetic anisotropy energy is mostly localized at the interface. However, only a slight to moderate decrease of PMA was reported in case of under-oxidized case when the oxygen is removed from the interface~\cite{MgOPMA} suggesting that other contributions exist besides the hybridization of orbitals between Fe and O orbitals. Therefore, the origin of the large PMA in Fe/MgO seems to be more complex.

In this letter, we investigate the PMA evolution as a function of the Fe and MgO layer thicknesses and different interfacial conditions using first principles calculations. In order to elucidate the microscopic mechanisms of PMA, we employ on-site projected analysis of PMA which enables identification not only of each layer's contribution to the total PMA value but also from states with different orbital characters. It is then illustrated that the PMA energy is not localized at the interface of Fe/MgO but rather distributed into the bulk with a damped oscillatory behavior as a function of distance from the interface. By analyzing interfacial and bulk contributions into the total PMA value, we conclude that the PMA has a more complex origin. The Fe-O bonding picture is an over-simplification of the anisotropy mechanism. It only contributes by a fraction of the total PMA value. Furthermore, we find that in most situations, states with $d_{yz(xz)}$ and $d_{z^2}$ character tend always to positively contribute to the PMA while those with $d_{xy}$ and $d_{x^2-y^2}$ character tend to favor in-plane anisotropy. Moreover, while MgO thickness has no influence on PMA, calculated perpendicular magnetic anisotropy is found to oscillate as a function of Fe thickness with a period of 2 monolayers (ML) and reaches a maximum value of 3.6~mJ/m$^2$.
\begin{figure}[t]
    \includegraphics[bb= 50 0 750 520,width=0.45\textwidth]{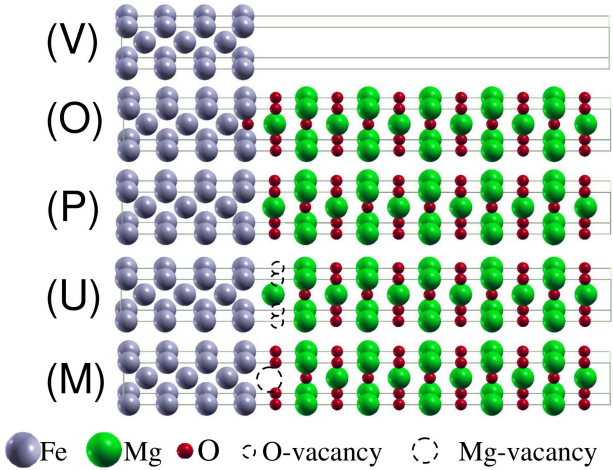}
   \caption{(Color online) Schematics of the calculated crystalline structures
for (V) Fe/Vacuum, (O) overoxidized Fe$_7$/MgO$_{11}$, (P) pure Fe$_7$/MgO$_{11}$, (U) underoxidized Fe$_7$/MgO$_{11}$, and (M)
Mg-vacancy in Fe$_7$/MgO$_{11}$. Fe, Mg and O are represented by silver, green and red balls respectively.}\label{fig1}
\end{figure}

Our first-principles calculations are based on density functional theory (DFT)
as implemented in the Vienna $ab ~initio$ simulation package (VASP)~\cite{vasp} 
within the framework of the projector augmented wave (PAW) potentials~\cite{paw} 
to describe electron-ion interaction and generalized gradient approximation 
(GGA)~\cite{gga} for exchange-correlation interactions. The calculations were
performed in three steps. First, full structural relaxations in shape and volume were performed until 
the forces become smaller than 0.001 eV/\AA ~for determining the most stable 
interfacial geometries. Next, the Kohn-Sham equations were solved with no spin-orbit
interaction taken into account to determine the ground state charge distribution of 
the system. Finally, the spin-orbit coupling was included and the total energy of 
the system was calculated as a function of the magnetization orientation. 
A 19$\times$19$\times$3 K-point mesh was used in our calculations. A plane wave 
energy cut-off equal to 520~eV for all calculations was used and is found to be
sufficient for our system. 

We use 11~ML of MgO and 7 ML of Fe for all 
the five structures considered as shown in Fig.~\ref{fig1}: (V) Fe$_7$/vacuum, (O) 
over-oxidized interface (with O inserted at the interfacial magnetic layer), 
(P) "pure" (O-terminated) interface, (U) under-oxidized (Mg-terminated) interface for investigation 
of oxidation conditions effects, and (M) Mg-vacancy at the interface. 
We introduce the orbital and layer resolved magnetic anisotropy as \mbox{$MA_{O,L}=\frac{-1}{a^2}(E^\perp_{O,L}-E^\parallel_{O,L})$}
where $a$ is the in-plane lattice constant and $E^{\perp(\parallel)}_{O,L}$ represents the energy contribution into anisotropy from 
layer $L$ and orbital $O$ for out-of-plane(in-plane) 
magnetization orientation in respect to the Fe/MgO interface. Positive values correspond to out-of-plane anisotropy. Taking into account 
that the magnetocrystalline anisotropy of the bulk iron is negligible, the interfacial magnetocrystalline anisotropy $K_S$ for the whole structure 
can be defined as \mbox{$K_S=\sum\limits_{O,L} MA_{O,L}$} where the sum is taken over all orbitals and layers of the Fe. 
Systematic calculations with Fe (respectively MgO) thickness varied between 5 and 13~ML with a fixed 11~ML of MgO (respectively Fe) for the odd number of monolayers. For
even number of MLs Fe thickness was varied between 6 and 12 MLs with a fixed 10ML thickness of MgO. 

\begin{figure}[t]
  \includegraphics[bb= 100 0 650 520,width=0.40\textwidth]{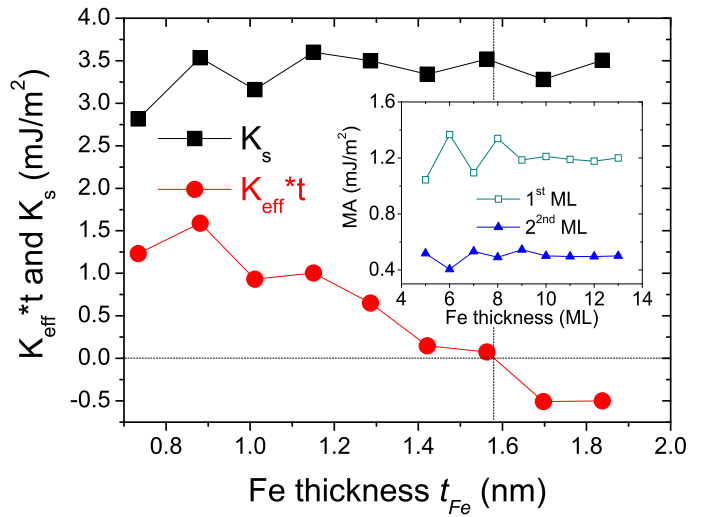}\\
   \caption{(Color online) Dependences of effective anisotropy and surface
anisotropy on Fe thickness 
   for Fe/MgO/Fe magnetic tunnel junctions, where MgO thickness is fixed to 11~ML
   and  Fe varies from 5 to 13 atomic layers. (inset) On-site projected magnetic anisotropy for the first and 
   second layer away from the interface as function of Fe thickness.}\label{fig2}
\end{figure}

In Fig. \ref{fig2} the surface anisotropy $K_S$ shows an oscillatory behavior as a function of Fe thickness $t_{Fe}$ with a period of 2~ML. 
The amplitude of $K_S$ increases linearly for odd number of layers and reaches a maximum of 3.5 mJ/m$^2$ at 9~ML of Fe thickness. 
At the same time for even number, it reaches a maximum value of 3.6~mJ/m$^2$ at 8~ML of Fe and then decreases for higher thicknesses.
Fig. \ref{fig2} also shows the effective anisotropy $K_{eff}\cdot t_{Fe}$ dependence on the thickness of Fe where $K_{eff}$ is defined as ${K_{eff}=K_S/t_{Fe}-2\pi M_s^2},$ the second term representing the demagnetizing energy which always favor in-plane anisotropy ($M_s$ representing the saturation magnetization of Fe layer). Since $K_S$ and $K_{eff}$ are related, $K_{eff}\cdot t_{Fe}$ also shows the same oscillatory behavior. However, $K_{eff}\cdot t_{Fe}$ decreases and reaches the cross point around 16~\AA ~due to the fact that the demagnetizing energy increases with the Fe thickness in agreement with recent experimental observations~\AA~\cite{keff}.
\begin{figure}[t]
  \includegraphics [bb=150 50 650 520,width=0.35\textwidth]{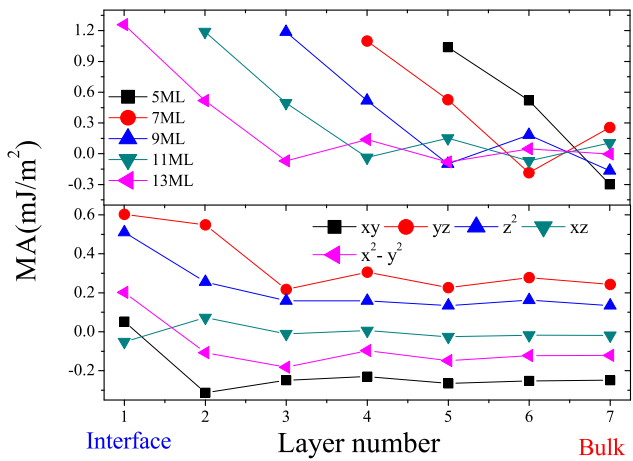}\\
   \caption{(Color online) (Top panel) On-site projected magnetic anisotropy for different Fe thicknesses in Fe/MgO system.  The curves are shifted with respect to each other for clarity. 
   MgO thikness is fixed at 11~ML. (Bottom panel) $d$-orbital resolved contribution to the magnetic anisotropy as function of layer number for Fe$_{13}$[MgO$_{11}$.}\label{fig3}
\end{figure}

In order to understand the origin of PMA behavior as a function of Fe thickness, we investigated the on-site projected
magnetic anisotropy for different thicknesses shown in Fig~\ref{fig3}(top panel).
One can see that the main contribution is localized at the first interfacial Fe layer and this value increases slightly as we
increase the Fe thickness and reaches a maximum of 1.2~mJ/m$^2$. At the same time, for even number of layers, the main contribution which is also
localized at the interface decreases as a function of Fe thickness from 1.4~mJ/m$^2$ to 1.2~mJ/$^2$ (inset of Fig~\ref{fig2}).
This value (1.2~mJ/$^2$) is consistent with that obtained by K.~Nakamura $et~al$ where one Fe atomic layer was used in calculations~\cite{Nakamura}.  
The second largest contribution to PMA comes from the second layer away from the interface. This contribution 
shows also 2 ML period of oscillations around 0.5 mJ/m$^2$  as a function of Fe thickness (inset Fig~\ref{fig2}). 

More generally, it is observed that although the PMA takes its origin at the Fe$_7$/MgO interface, significant contributions to the PMA energy come from the bulk Fe layers, these contributions exhibiting an attenuated oscillatory behavior as a function of distance to the interface~\cite{Samsung}. The observation of such oscillatory behavior has been 
recently reported experimentally in different systems~\cite{qws, qws1}. These oscillations were attributed to quantum well 
oscillations in a minority-spin $d$-band at the Fermi level. As one can see from Fig.~\ref{fig3}(bottom panel) where orbital resolved contributions to the magnetic anisotropy are shown, 
$\Delta_5$ (d$_{xz,yz}$) orbitals present in Fe around Fermi level for minority electrons, are dominating in PMA and seem to be at the origin of these 2 ML oscillations.

\begin{table} [t]
\caption {Surface anisotropy $K_S$ in mJ/m$^2$ for all the considered structures and corresponding contributions of the first two layers next to the interface.
The  $K_S$ values are given for 2 interfaces. The values for Fe$_5$[MgO]$_3$ are taken
from Ref.~\cite{MgOPMA}}
\label{Table1}
            \begin{tabular}{llllll}
                        \hline\hline
          Structure          &PMA:~~~      &~~~~~ ~~$K_S$       &~~~~$MA_{o,1}+MA_{o,2}$   \\                            
        \hline
         Fe$_7$[MgO]$_{11}$  & pure        &~~~~~ ~~~3.15       &~~~~~~~~ ~~~1.63        \\
              ~~(MTJ)        & under-      &~~~~~ ~~~2.84       &~~~~~~~~ ~~~1.22        \\
                             & over-       &~~~~~ ~~~0.25       &~~~~~~~~ ~~-1.56        \\
                         & Mg-Vacancy      &~~~~~ ~~~3.15       &~~~~~~~~ ~~~1.39        \\
             \hline
          Fe$_7$[MgO]$_{11}$  & pure       &~~~~~ ~~~2.70       &~~~~~~~~ ~~~1.82        \\
              ~~(Vacuum)      & under-     &~~~~~ ~~~2.08       &~~~~~~~~ ~~~1.17        \\
                              & over-      &~~~~~ ~~-0.33       &~~~~~~~~ ~~-1.45        \\
              \hline          
             Fe$_5$[MgO]$_3$  & pure       &~~~~~ ~~~2.93       &~~~~~~~~ ~~~1.56        \\
              ~~(MTJ)         & under-     &~~~~~ ~~~2.27       &~~~~~~~~ ~~~0.89        \\
                              & over-      &~~~~~ ~~~0.98       &~~~~~~~~ ~~-1.22        \\
              \hline        
            Fe$_7$[MgO]$_1$   & pure       &~~~~~ ~~~2.62       &~~~~~~~~ ~~~1.85        \\
             ~~(Vacuum)       & under-     &~~~~~ ~~~2.11       &~~~~~~~~ ~~~1.17        \\
                          & Mg-Vacancy     &~~~~~ ~~~1.20       &~~~~~~~~ ~~~0.65        \\
             \hline 
          Fe$_7$ (Vacuum)    &            &~~~~~ ~~~1.74       &~~~~~~~~ ~~~0.80        \\
                                 
        \hline\hline
   \end{tabular}
\end{table}

So far, we discussed the case of pure interface represented in Fig.~\ref{fig1}(P). We now present results of the impact of interfacial conditions for different configurations and structures rerpresented in Fig.~\ref{fig1}(V,U,O,M). In Table~\ref{Table1} we summarize $K_S$ and $MA_{o,1}+MA_{o,2}$ values since the latter represent the main contribution to $K_S$ as explained above. The calculations show that PMA reaches its maximum in the case of pure interfaces for all considered structures. The PMA is slightly reduced compared to the pure case for under-oxidized interfaces and strongly reduced for over-oxidized interfaces. These results are in agreement with our previous report~\cite{MgOPMA} and 
recent experiments~\cite{LaviniaIEEE,APL2012}. For slab structures comprising both Fe/MgO and Fe/Vacuum interfaces 
(Fe(001) surface), we obtained 2.7~mJ/m$^2$ for Fe$_7$[MgO]$_{11}$ which is in good agreement with previous report of 2.85 mJ/m$^2$ for Fe$_9$[MgO]$_9$ slab 
\cite{TsymbalPMA}. Furthermore, we also investigated the PMA dependence on MgO thickness and found that $K_S$ is not affected by it. 
The PMA variation versus MgO thickness for all cases is less than $\pm$0.1~mJ/m$^2$ which is in a good agreement with experiments~\cite{thickMgO}. For example, 1~ML of MgO on top of 7~MLs of Fe gives rise to PMA of 2.62~mJ/m$^2$ which is comparable with case of 11~MLs of MgO as seen in Table~\ref{Table1}. Interestingly, Table~\ref{Table1} shows that while the PMA is relatively insensitive to MgO thickness in the pure and underoxidized cases, this is no longer the case when Mg vacancies are present. The latter significantly affect the PMA for thin MgO layers as seen for Fe$_7$/MgO$_1$ structure.  

\begin{figure} [t]
  \includegraphics[width=0.45\textwidth]{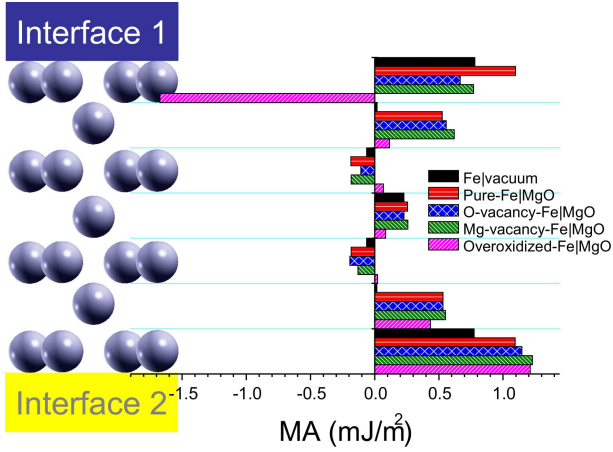}\\
   \caption{(Color online) On-site projected magnetic anisotropy for different interface1/Fe$_7$/interface2 structures. Interface1:  Fe$_7$/vacuum, pure Fe$_7$[MgO]$_{11}$, under-oxidized Fe$_7$[MgO]${_11}$(O-vacancy), Mg-vacany Fe$_7$/MgO and over-oxidized  Fe$_7$[MgO]$_{11}$. Interface2: Fe$_7$/vacuum and pure Fe$_7$/MgO}\label{fig4}
\end{figure}

\begin{figure} [b]
  \includegraphics[bb=150 50 650 520,width=0.35\textwidth]{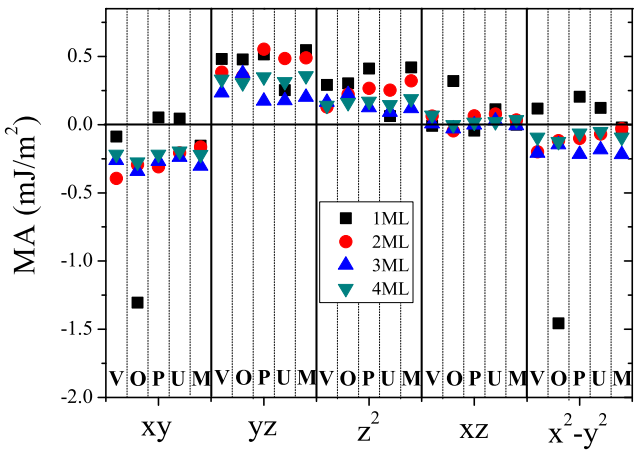}\\
   \caption{(Color online) d-orbital resolved contribution to the magnetic anisotropy for different interfacial conditions.
    V, O, P, U and M letters correspond to Fe$_7$/Vaccum, overoxidized Fe$_7$/MgO, pure Fe$_7$/MgO, underoxidized Fe$_7$/MgO and
     Fe$_7$/MgO with Mg vacancy case, respectively.}\label{fig5}
\end{figure}
In order to elucidate the influence of interfacial conditions on magnetic anisotropy, we again use on-site projected MA. In Figure~\ref{fig4} we summarize MA contributions from each individual Fe layers sandwiched between interface 1 considering the various investigated cases(see Fig.~\ref{fig1}) and interface 2 being kept either Fe/Vacuum or pure Fe/MgO. By analyzing data presented in Fig.~\ref{fig4} one can identify several mechanisms of the interfacial PMA and separate them in three main origins~\cite{Samsung2}. The first one originates from the symmetry breaking at the Fe/vacuum interface case and one can see that the main contribution to the total PMA of pure Fe slab (0.87~mJ/m$^2$, see Table~\ref{Table1}) is located at the first layer and is equal to 0.78~mJ/m$^2$ (Fig.~\ref{fig4}(black). The second one is due to hybridizations between the transition metal and the insulator orbitals at the interface, mainly in the first layer. This mechanism is active in case of pure interface and provides the largest total PMA through its contribution from the first Fe layer (Fig.~\ref{fig4}(red)). Finally, the third one is localized in the second layer from the interface and results from the hybridizations between the Fe-$3d$ orbitals of the first and second layers since the presence of Mg or O vacancies have little effect on this contribution as indicated by the relative close amplitudes of the blue, green and red bars in Fig.~\ref{fig4}. Instead, these vacancies result in decreasing slightly the total interfacial PMA by reducing the contribution from the first layer by about between one-half and one-third compared to the pure case. The most dramatic impact on interfacial PMA is in case of overoxidized interface as seen in Table~\ref{Table1}. The reason is that when oxygen is added to interface 1 in Fe$_7$/MgO$_{11}$ MTJ (Fig.~\ref{fig1}(O)), a significant abrupt change in PMA occurs where the first layer contribution becomes negative with a value of -1.7~mJ/m$^2$ (Fig.~\ref{fig4}(magenta)). This could be explained by strong in-plane hybridization between Fe-$3d$ and O-$2p$ orbitals in the FeO layer. Indeed, by looking at the orbital-resolved contribution to the anisotropy shown in Fig.~\ref{fig5}, one can see clearly that in case of over-oxidation (O) the contribution of in-plane orbitals (d$_{xy}$ and d$_{x^2-y^2}$) becomes strongly negative in the first layer (black squares) compared to all other cases (V,P,U,M). In fact, analysis of the orbital contribution to the MA shown in Fig.~\ref{fig5} allows to clarify and elucidate even further the microscopic origin of the PMA in Fe/MgO. First, we can see that in general the out-of-plane orbitals (d$_{z^2}$ and d$_{xz,yz}$) always try to align the magnetization out-of-plane while the in-plane orbitals (d$_{xy}$ and d$_{x^2-y^2}$) have a tendency to align the magnetization in-plane. Next, the contributions from out-of-plane and in-plane orbitals in the bulk of the layer compensate each other giving rise to a negligible MA. However, at the interface, more precisely in the first layer, the contribution of the out-of-plane orbitals increases and dominates except for the underoxidized case (U). This concerns especially the contribution from d$_{z^2}$ due to the hybridization with O orbitals in case of pure (P) and Mg vacancy (M). Interestingly, the in-plane orbitals contribution (d$_{xy}$ and d$_{x^2-y^2}$) in the first layer in case of pure and underoxidized interface also tend to align the magnetization in out-of-plane direction with smaller values compared to d$_{z^2}$ and d$_{xz,yz}$. Finally, the presence of Mg atom at the interface seems to be crucial in changing the in-plane contribution sign from negative to positive. This becomes clear by comparing the pure case with the O and Mg vacancies cases. One can see that while the oxygen vacancies have strong influence on out-of-plane orbital contributions, the Mg vacancies seem to affect only the in-plane ones. This could be related to the fact that oxygen atom is located on top of the Fe atom while Mg is located in the hollow site.

In conclusion, using first-principles calculations, we unveiled the microscopic mechanisms of PMA by evaluating the orbital and on-site projected contributions to magnetic anisotropy in Fe/MgO interfaces and MTJs with different interfacial conditions. Our results indicate that the origin of the large PMA oberved in MgO-based MTJ is more complex and much richer than described so far by only considering the hybridization between Fe-$3d$ and O-$2p$ orbitals. Furthermore, we demonstrated that the PMA energy is not localized at the interface but is distributed also within the bulk of the Fe layers showing a damped oscillatory character as a function of Fe thickness and distance to the interface with a period of 2~ML. The PMA reaches a maximum value of 3.6~mJ/m$^2$ for two interfaces. This oscillatory character is due to the confinement of minority electrons in d$_{xz,yz}$ orbitals inside the Fe film and between the Mg barrier. It is also found in most situations that states with $d_{yz(xz)}$ and $d_{z^2}$ character tend always to favour PMA while those with $d_{xy}$ and $d_{x^2-y^2}$ character tend to favor in-plane anisotropy. We expect similar mechanisms may be found in other metal/insulator structures.

We acknowledge D.~Apalkov, O.~Mryasov, W.~H.~Butler and A.~Smogunov for fruitful discussions.


\begin{thebibliography}{200}
\bibitem{Samsung2} Two of these mechanisms have been also pointed in Ref.~\cite{Samsung}.
\end{thebibliography}
\end{document}